\documentclass[a4paper,usenatbib,fleqn,useAMS]{mnras}
\usepackage{newtxtext,newtxmath}
\usepackage[T1]{fontenc}
\usepackage{ae,aecompl}

\usepackage{hyperref}
\usepackage{graphicx}

\begin{document}
\title{The mass of the black hole in 1A~0620--00, revisiting the ellipsoidal light curve modeling}
\author[T. F. J. van Grunsven et al.]
{Theo F.J. van Grunsven,$^{1,2}$ \thanks{Email: t.f.j.van.grunsven@sron.nl}
Peter G. Jonker,$^{1,2}$
Frank Verbunt$^{2}$
\newauthor
and Edward L. Robinson$^{3}$\\
\\
$^1$SRON Netherlands Institute for Space Research, Sorbonnelaan 2, 3584 CA Utrecht, The Netherlands\\
$^2$Department of Astrophysics/IMAPP, Radboud University Nijmegen, P.O. Box 9010, 6500 GL Nijmegen, The Netherlands\\
$^3$Department of Astronomy, University of Texas, 1 University Station, C1400, Austin, Texas 78712, USA}
\maketitle

\begin{abstract}
The mass distribution of stellar mass black holes can provide important clues to supernova modeling, but observationally it is still ill constrained. Therefore it is of importance to make black hole mass measurements as accurate as possible. The X-ray transient 1A~0620--00 is well studied, with a published black hole mass of $6.61\pm0.25\,$M$_{\sun}$, based on an orbital inclination $i$ of $51\fdg0\pm0\fdg9$. This was obtained by \cite{2010ApJ...710.1127C}, as an average of independent fits to $V$-, $I$- and $H$-band light curves. In this work we perform an independent check on the value of $i$ by re-analyzing existing YALO/SMARTS $V$-, $I$- and $H$-band photometry, using different modeling software and fitting strategy. Performing a fit to the three light curves simultaneously, we obtain a value for $i$ of $54\fdg1\pm1\fdg1$, resulting in a black hole mass of  $5.86\pm0.24\,$M$_{\sun}$. Applying the same model to the light  curves individually, we obtain $58\fdg2\pm1\fdg9$, $53\fdg6\pm1\fdg6$ and $50\fdg5\pm2\fdg2$ for $V$-, $I$- and $H$-band, respectively, where the differences in best-fitting  $i$ are caused by  the contribution of the residual accretion disc light in the three diferent bands. We conclude that the mass determination of this black hole may still be subject to systematic effects exceeding the statistical uncertainty. Obtaining more accurate masses would be greatly helped by continuous phase-resolved spectroscopic observations simultaneous with photometry.   
 \end{abstract}


\section{INTRODUCTION}
It is thought that stellar-mass black holes (hereafter: BHs) are 
formed through the collapse
of the core of a high-mass star. However, the relation between the spin, mass and natal kick of the newly formed BH and the mass, evolutionary history, and parameters such as metallicity and spin of
the progenitor star, are strongly model dependent.
\citep[e.g.][]{2012ApJ...749...91F}.
The observationally determined distribution of the masses of BHs, 
including the dearth 
of BHs with masses in the range of 2--5 M$_{\sun}$ 
\citep{2010ApJ...725.1918O, 2011ApJ...741..103F}
provides powerful constraints on the supernova models 
\citep[e.g.][]{2012ApJ...757...69U}.
Currently the mass is known of some 20 stellar-mass BHs. All these BHs 
are in binaries,
and most of them are soft X-ray transients, with a low-mass donor star
\citep[see the review by][]{2014SSRv..183..223C}.

To determine the mass of a BH in such a binary, standard techniques for 
the analysis of a binary
with two extended stars are adapted by the replacement of one extended 
star with a compact star
surrounded by an accretion disc.
Furthermore, in the case of soft X-ray transients it is usually assumed 
that the mass donor corotates
with the orbit and fills its Roche lobe \citep[e.g.][]{2001icbs.book.....H}.  In that case,
the observed amplitude of the radial-velocity curve of the donor, combined with 
the width of the
rotationally-broadened spectral lines gives the mass ratio between the 
donor and the compact star.
The ellipsoidal variation of the observed flux from the donor then allows 
determination of the orbital inclination.
A complicating factor in this is the extra flux from the accretion disc 
and from the hot spot caused
by the impact of the mass stream from the donor on (the outer edge of) 
the accretion disc.
The orbital variation of this extra flux depends on the temperature 
distribution across disc and
hot spot, and on their geometrical structure, all of which are virtually 
unknown.
This flux must be subtracted from the observed flux in order to 
reconstruct the correct amplitude
and phase dependence of the pure ellipsoidal variation.

The flux from the accretion disc can be determined when spectral
lines of the donor star are detected in the spectrum of the 
source.  The
observed flux $F_\mathrm{o}$ is the sum of the flux of the donor $F_2$ 
and the flux of the
disc $F_\mathrm{d}$. By subtracting trial template spectra one 
determines for which spectral
type and flux level $F_2$ the absorption lines disappear from the remaining 
spectrum, and thereby determines
$F_2$ and $F_\mathrm{d}=F_\mathrm{o}-F_2$ as a function of wavelength. 
This method assumes
that the disc does not contribute to the absorption lines, i.e.\ that the disc 
spectrum is smooth.
The method may be repeated for each orbital phase. If a spectrum is 
available at one orbital phase
only, the flux from the donor at other phases can be determined for 
given inclination and mass ratio
from a model for the ellipsoidal variation.

\cite*{2008MNRAS.384..849N}  apply the method to the orbital average 
spectrum
of 1A~0620--00 (V616 Mon, hereafter A0620), a well-studied BH binary system, thus determining the orbital average of the disc contibution 
(see their fig.\,1).
In an effort to determine the ellipsoidal variation they assume that 
the broad H\,$\alpha$  emission line flux and the disc continuum
flux are constant with orbital phase, and thereby convert the orbital 
variations
of the equivalent width  into the orbital variation of the flux from the 
donor. The
orbital variations thus found are incompatible with ellipsoidal 
variation, and the authors conclude
that disc and/or line flux in fact do vary with orbital phase, a conclusion also reached by \citet{2015ApJ...808...80C,2016ApJ...822...99C}.

Several inclination determinations of A0620 are listed by \cite{2014SSRv..183..223C}. They range from $36\fdg7$ \citep*{1994MNRAS.268..756S} to $38\degr$-$75\degr$ \citep{2001AJ....121.2212F}. The most recent value is $51\fdg0\pm0\fdg9$ (\citealt*{2010ApJ...710.1127C}, henceforth CBO10).
We reanalyse the $V$, $I$ and $H$ filter-band YALO/SMARTS light curves used by CBO10 (3 out of their 12 different light curves obtained at different observatories, see table 3 in their article) to investigate the dependence of the 
results -- most importantly the orbital inclination and the derived mass of the black hole -- on the methods used. In particular, we use a different computer code, we fit the 3 light curves simultaneously, and we add a hot spot on the outer edge of the disc to our light curve model.

\section{DATA}
\label{sec:data}

\begin{figure}
  \includegraphics[width=80 mm]{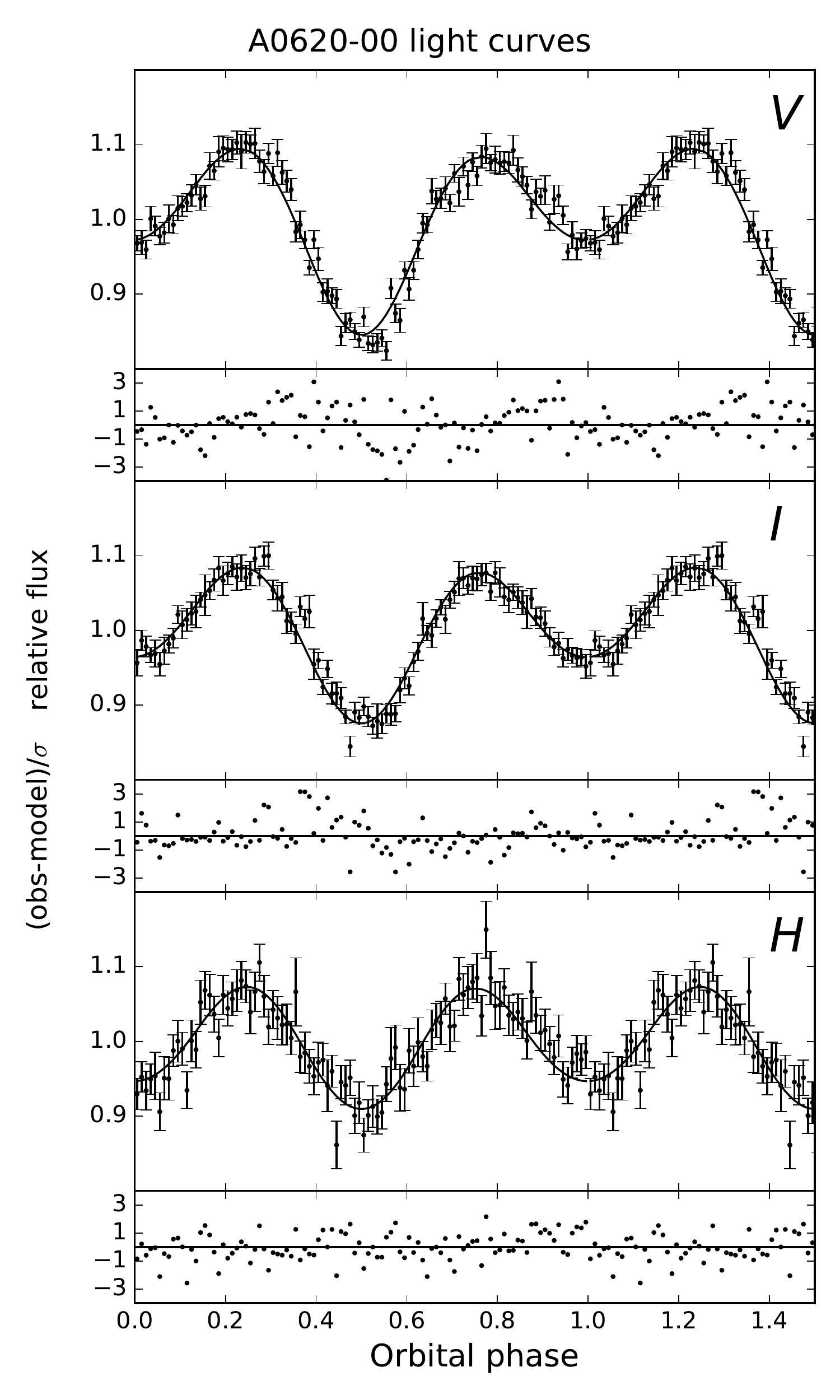}
  \caption{A0620 folded light curves. The data are shown binned in 100 equal phase intervals, but the fitting is done to the unbinned data. We plot 1.5 periods for clarity. The light curves are normalized so that the average flux is equal to 1.0 for each photometric filter. The solid lines represent the best fit to the three light curves simultaneously, without phase shift (see Section~\ref{sec:discussion}). The residuals have been weighted with the observational uncertainties.} 
 \label{fig:1}
\end{figure}

This work is based on the YALO/SMARTS dataset used by CBO10. Typically one observation per night was taken. The ANDICAM instrument \citep{2003SPIE.4841..827D} was used first on the YALO 1\,m telescope, and on the SMARTS 1.3\,m telescope from February 2003 onwards.  One observation consisted of one \emph{V} and one \emph{I} exposure with simultaneous dithered \emph{H} exposures. $V$ and $I$ exposure times were 11 minutes each on the 1\,m and 6 minutes on the 1.3\,m telescope. On the 1\,m, seven dithered 90\,s $H$-band images were taken during each $V$ or $I$ exposure, on the 1.3\,m eight dithered 30\,s $H$-band images. 
The $H$-band dithered images were combined in sets of 5, 5, and 4 per \emph{V,I} exposure pair on the 1\,m telescope, and in 2 sets of 8 on the 1.3\,m.  Deviations from this rule occur because a number of images was discarded for quality reasons. 
CBO10 also discarded large sections of the full dataset where the disc was in an "active" state. The resulting light curves represent a state with low disc activity to which the system returned repeatedly over the course of four years.

Of the remaining data we had to discard ten $H$-band data points because the FITS headers showed they were derived by combining sub-exposures taken on different nights. The resulting number of data points equals 750, 732 and 852 in \emph{V-, I-} and \emph{H}-band, respectively (our dataset is available as Supplementary material). In our analysis $H$-band timestamps reflect the HJD midpoint of the combined sequence. The statistical errors of the differential photometry are reported by \cite{2008ApJ...673L.159C} as 0.04\,mag in \emph{V} and \emph{I}, and 0.08\,mag in \emph{H}. These can only be roughly valid in an average sense; the uncertainties of individual observations will vary as a result of varying observing conditions but also because the total exposure times of combined $H$-band images varies, and because the signal to noise ratio differs between exposures taken with the 1\,m and 1.3\,m telescopes. Figure~\ref{fig:1} shows the folded and binned light curves, where we used the spectroscopic ephemeris of \cite{2010A&A...516A..58G}.

Additional information is derived from spectroscopic observations: for the fractions of light contributed by the accretion disc in each band we use the values resulting from the analysis in CBO10, given at phase 0.554 (corresponding to 0.054 in our phase convention, as the $T_0$ in CBO10 corresponds to upper conjunction of the companion star, whereas our $T_0$ corresponds to the upper conjunction of the BH), which are: $0.35\pm0.03$, $0.25\pm 0.03$ and $0.13\pm0.02$ in $V$-, $I$- and $H$-band, respectively; we treat these values as extra data points when calculating the goodness-of-fit measure $\chi^2$. For the ratio $q$ between the mass of the secondary star to the BH mass we adopt $0.060\pm{0.004}$ \citep*{2008MNRAS.384..849N} as the \emph{prior} distribution (see Section \ref{sec:modeling}). 

\section{Modeling the light curve}
\label{sec:modeling}
We model the light curve of A0620 using the (slightly modified) 
\textsc{XRbinary} program written by E.L. Robinson 
(e.g.~\citealt{2010ApJ...709..251B})\footnote{a full description of the program is available at \\ 
\url{http://www.as.utexas.edu/~elr/Robinson/XRbinary.pdf}}.
The program calculates the model flux at each orbital phase. To do so, 
it assumes a binary system consisting of a compact primary object 
surrounded by an accretion disc that is fed by mass transfer from a 
secondary star that fills its Roche lobe. The secondary is assumed to 
move in a circular orbit, and to corotate, so that the Roche geometry 
applies.
The variation of the effective temperature $T_\mathrm{eff}$ over the 
stellar surface is described by gravity darkening, $T_\mathrm{eff}\propto 
|g|^\beta$, where $g$ is the surface gravity 
\citep{1924MNRAS..84..665V}.
The exponent $\beta$ depends only on $T_\mathrm{eff}$ and is taken from  a 
table based on \cite{2000A&A...359..289C}.
The flux in the $V$, $I$ and $H$ filters is computed for each surface 
element from Kurucz
stellar atmosphere models with $3500 < T_\mathrm{eff}(\mathrm{K})<8000$, with 
a 4-parameter
limb-darkening law \citep{2000A&A...363.1081C}.

We assume a cylindrically symmetric accretion disc, with inner radius 
$r_\mathrm{in}$ and outer radius $r_\mathrm{out}$.
The \mbox{semi-height} of the disc is given by
\begin{equation}
h(r) = \left(\frac{r-r_\mathrm{in}}{r_\mathrm{out}-r_\mathrm{in}}\right)^nh_\mathrm{out}\quad \mathrm{for}\enspace
   r_\mathrm{in}\le r \le r_\mathrm{out} 
\label{e:height}
\end{equation}
The surface elements of the disc are assumed to emit a blackbody 
spectrum -- not limb darkened -- with temperature
\begin{equation}
T(r)=T_\mathrm{out} \left(\frac{r}{r_\mathrm{out}}\right)^\gamma\quad \mathrm{for}\enspace
   r_\mathrm{in}\le r \le r_\mathrm{out}
\label{e:Tdisc}
\end{equation}
The side of the disc is a cylindrical surface with a single bright spot over the full height $2h_\mathrm{out}$, with center position $\zeta_\mathrm{spot}$ and full width $\Delta\zeta_\mathrm{spot}$. The combined temperature profile is
\begin{gather}
 \begin{split}
T(\zeta) = T_\mathrm{edge} + &(T_\mathrm{spot}-T_\mathrm{edge})
\cos^2\left[\frac{\pi\left(\zeta-\zeta_\mathrm{spot}\right)}{\Delta\zeta_\mathrm{spot}}\right]\\
   &\text{for}\;|\zeta-\zeta_\mathrm{spot}|<{\Delta\zeta_\mathrm{spot}/2}
\end{split}\\
T(\zeta)=T_\mathrm{edge}\;
 \text{for}\;|\zeta-\zeta_\mathrm{spot}|>\Delta\zeta_\mathrm{spot}/2
\end{gather}
The angle $\zeta$ is defined such that $\zeta$ is zero on the extension of the (corotating) line connecting the center of mass of the companion star to that of the BH; therefore a spot with $\zeta_\mathrm{spot}$ equal to zero is maximally visible at the upper conjunction of the companion. $\zeta$ increases counter to the direction of the orbital motion of the companion.

We use 10\,000 surface elements of roughly the same area for the secondary, and 10\,000 for the disc 
surfaces and edge.
At each orbital phase the angle between each of the  surface elements 
and the direction to Earth
is computed, and the flux in the direction of Earth is computed from the 
effective temperature
and (where applicable) limb darkening. It is checked which surface 
elements are occulted
by any of the other surfaces, and the fluxes of the unocculted elements 
are added. 
The model provides the flux of the star $f_\mathrm{s}$ and of the
accretion disc $f_\mathrm{d}$, and thereby the total flux $f = f_\mathrm{s}+f_\mathrm{d}$
as well as the fraction $f_\mathrm{d}/f$ of the total flux contributed
by the disc in $V$, $I$ and $H$, at each orbital phase.

\begin{table}
\caption{Fit parameters. Distances are expressed in units of the orbital separation $a$; $r_\mathrm{L}$ is the distance to the inner Lagrangian point.}
\label{table:parameters}
\begin{tabular}{lcl}
\multicolumn{3}{c}{\bf Basic model, fixed parameters} \\
parameter  & value &   \\
\hline
$T_\mathrm{eff}$(K) & 4600 & average temperature donor, ref. [1] \\
4$\beta$ & 0.415 & coefficient gravity darkening, ref. [2]\\
$r_\mathrm{in}$ & 0.10 & inner radius of the disc, Eq. \ref{e:height}  \\
$n$ & 1.2 & exponent disc flaring, Eq. \ref{e:height}\\
$\gamma$ & -0.75 & exponent disc temperature, Eq. \ref{e:Tdisc} \\
$\Delta\phi$ & 0.0 & phase offset\\
\hline
\end{tabular}
References: 
  [1] {\cite*{2001AJ....122.2668G}};  [2] {\cite{2000A&A...359..289C}}\\
\\
\begin{tabular}{p{5mm}cl }
\multicolumn{3}{c}{\bf Basic model, fitted parameters and priors}\\
parameter & prior &   \\
\hline
$q$ & Gaussian, $q=0.060\pm0.004$ & mass ratio, ref. [3]\\
$i$ & $P(i) = \sin(\frac{\pi i}{180}$); $0\degr\leq i\leq90\degr$ & inclination \\
$r_\mathrm{out} $ &  $r_\mathrm{in}+0.02\leq r_\mathrm{out}\leq 0.9r_L$ &
outer radius of the disc\\
$h_\mathrm{out}$ & $0.01r_\mathrm{out}\leq h_\mathrm{out}\leq r_\mathrm{out}$
& outer disc half-height\\
$T_\mathrm{out}$ & $250\,\mathrm{K}\leq T_\mathrm{out} \leq 20\,000$\,K &
temperature outer disc \\
$T_\mathrm{edge}$ & $T_\mathrm{out}\leq T_\mathrm{edge} \leq 20\,000$\,K &
temperature disc edge \\
$T_\mathrm{spot}$ & $T_\mathrm{edge}\leq T_\mathrm{spot} \leq 20\,000$\,K &
temperature hot spot \\
$\zeta_\mathrm{spot}$ & $0\degr\leq\zeta_\mathrm{spot}\leq360\degr$ & phase of hot spot centre\\
$\Delta\zeta_\mathrm{spot}$ & $0\leq\Delta\zeta_\mathrm{spot}\leq180\degr$ &
full width of hot spot\\
\hline
\end{tabular}
Reference:  
  [3] {\cite*{2008MNRAS.384..849N}}\\
\begin{tabular}{lccl}\\
\multicolumn{4}{c}{\bf Variant models  (see Section \ref{sec:discussion})} \\
 & parameter & prior or fixed value\\
\hline
V1 & $T_\mathrm{eff}$(K) & 4200 \\
V2 & $r_\mathrm{in}$ & $0.02$ \\
V3 & $\gamma$ & 0.0\\
V4 & 4$\beta$ & 0.400\\
V5 & $T_\mathrm{edge}$ & $T_\mathrm{edge}=T_\mathrm{out}$\\
V6 & $\Delta\phi$ & $-0.1<\Delta\phi<0.1$  & orbital phase offset\\
\hline
\end{tabular}    
\end{table}

The parameters necessary to describe the model are listed in 
Table\,\ref{table:parameters}. All lengths are expressed as a fraction of the binary orbital separation $a$.
Most prior probability distributions for the fitted parameters (Section \ref{sec:fitting}) are flat over an allowed range. Outside these ranges their probability is zero. This prevents problems caused by unphysical parameter proposals in the Markov-Chain Monte Carlo procedure. The distribution for $i$ reflects the assumption that the \emph{a priori} orientation of the binary orbital plane is random.

Preliminary modeling showed that of the disc parameters $r_{in}$, $n$ and $\gamma$
 are very poorly constrained by the data. Therefore we assigned them fixed values. We repeat the fitting procedure with alternative values to ensure we do not introduce an appreciable systematic error by doing this, as described in Section~\ref{sec:discussion}.  For reasons also discussed in Section~\ref{sec:discussion} we try fitting the light curves while allowing a variable phase offset $\Delta\phi$ between the spectroscopic $T_0$ and the $T_0$ of the fitted light curve.

\section{LIGHT CURVE FITTING}
\label{sec:fitting}
For estimating the probability distributions of the model parameters (given the observations) we use a Markov-Chain Monte Carlo (MCMC) sampling method to find the probability distribution of free parameters of the model (9 for the basic model and all variants except V1, which has 10),  given the observed data, the fixed parameters, and (where available) {\em{a priori}} knowledge of the free parameters (so-called priors). According to Bayes' theorem:
\begin{equation}
P(\bmath{a}|\mathrm{D})\propto P(\mathrm{D}|\bmath{a})\,P(\bmath{a})
\end{equation}
where $\bmath{a} =a_1,a_2,\dots,a_\mathrm{M}$ is a realization of the M\nobreakdash-dimensional vector
of variable parameters, $P(\bmath{a}|\rm{D})$ is the probability of $\bmath{a}$ given the data $\mathrm{D}$, $P(\mathrm{D}|\bmath{a})$ is the probability of $\mathrm{D}$ for a given $\bmath{a}$, and $P(\bmath{a})$ is the \emph{a priori} probability of $\bmath{a}$. 
In our case $\mathrm{D}$ consists of the photometric data $(x_i,y_i)~i = 1,2,\dots,\mathrm{N}$; $x_i$ is the orbital phase of each flux measurement $y_i$. Additional data points are $(0.054,y_\mathrm{N+1})$, $(0.054,y_\mathrm{N+2})$\,and $(0.054,y_\mathrm{N+3})$, where $y_\mathrm{N+1} = 0.35$, $y_\mathrm{N+2}=0.25$ and $y_\mathrm{N+1}=0.13$ are the three spectroscopic disc fractions at phase $\phi = 0.054$. Each data point has an associated uncertainty, assumed to have a Gaussian distribution with width $\sigma_i$. \\
Assuming the \emph{a priori} distributions of the parameters are independent of each other $P(\bmath{a})$ can be written as the product $P(\bmath{a})= \prod_{i=1}^\mathrm{M} P(a_\mathrm{i}) $. These priors are listed in Table\,\ref{table:parameters}.\\
For each realization of $\bmath{a}$ we compute the model light curve values $m_i(x_i,\bmath{a})~i = 1,2,\dots,\mathrm{N}$, as well as $m_i(0.054,\bmath{a})$~$i=$ N+1,\,N+2,\,N+3, the disc fractions at phase 0.054. The probability of $\bmath{a}$ can be written (ignoring a constant)
\begin{gather}
\begin{split}
\ln P(\bmath{a}|\mathrm{D}) &=  \sum_{j=1}^{\mathrm{M}}{\ln P(\bmath{a}_j)}+ \sum_{i=1}^{\mathrm{N}+3}{\ln P(\mathrm{D}|m_i(x_i,\bmath{a}))}\\
 &= \sum_{j=1}^{\mathrm{M}}{\ln P(\bmath{a}_j)} - 
\sum_{i=1}^{\mathrm{N+3}}{\frac{(m_i(x_i,\bmath{a})-y_i)^2}{2\sigma_i^2}}\\
 & = \sum_{j=1}^{\mathrm{M}}{\ln P(\bmath{a}_j)} -\frac{ \chi^2(\mathbf{a})}{2}
\end{split}
\end{gather}

\cite{goodman2010ensemble} describe a family of ensemble samplers with affine invariance, the performance of which is unaffected by affine transformations of parameter space. The algorithm automatically takes care of generating parameter proposals that  efficiently sample their \emph{a posteriori} distribution. We use a Python implementation of the algorithm, emcee, already used in many astrophysics projects \citep{2013PASP..125..306F}. This exploits the inherent parallelism of the ensemble samplers to take advantage of multiple CPU cores without extra effort. The Simplified Wrapper and Interface Generator (SWIG, \url{www.swig.org})  is used to generate an interface between Python code and XRbinary, which is written in C.

For each parameter proposal $\ln P$ is calculated. Computing the model flux at the orbital phase of each of the 2331 data points would require a large amount of computing capacity. This computation can be reduced by binning the data, but we have chosen to use interpolation of the computed model light curves instead: the model flux is calculated at 36 evenly spaced phases, and its value at the phase of each datapoint is obtained using periodic cubic spline interpolation. This introduces negligible error as the light curves are smooth.

Assuming that the model is a good description of the system, and since the three SMARTS light curves cover (almost) the same period, fitting them with the same model should yield a single set of parameter values. We therefore determine the probability distributions of these by fitting all three light curves simultaneously. We perform an MCMC run consisting of 400 chains running in parallel. Each of the chains is initiated with a parameter vector drawn from a narrow Gaussian distribution around a value close to the expected peak probability, as determined from exploratory runs. First, the chains are run for 1,000 iterations of ``burn in" to allow them to become distributed reasonably close to the target probability density; these iterations are discarded. The main run consists of 10,000 iterations, so we calculate a total of 4,000,000 light curves and resulting probabilities. As convergence is not always sufficient at the beginning of the main run, we use only the last 5,000 iterations for deriving the parameter probability distributions.
Finally, we use the Gelman-Rubin diagnostic \citep{1992StaSc...7..457G} to verify convergence of the MCMC chains.

\section{RESULTS}
\label{sec:results}

\begin{figure*}
  \includegraphics[width = 140mm]{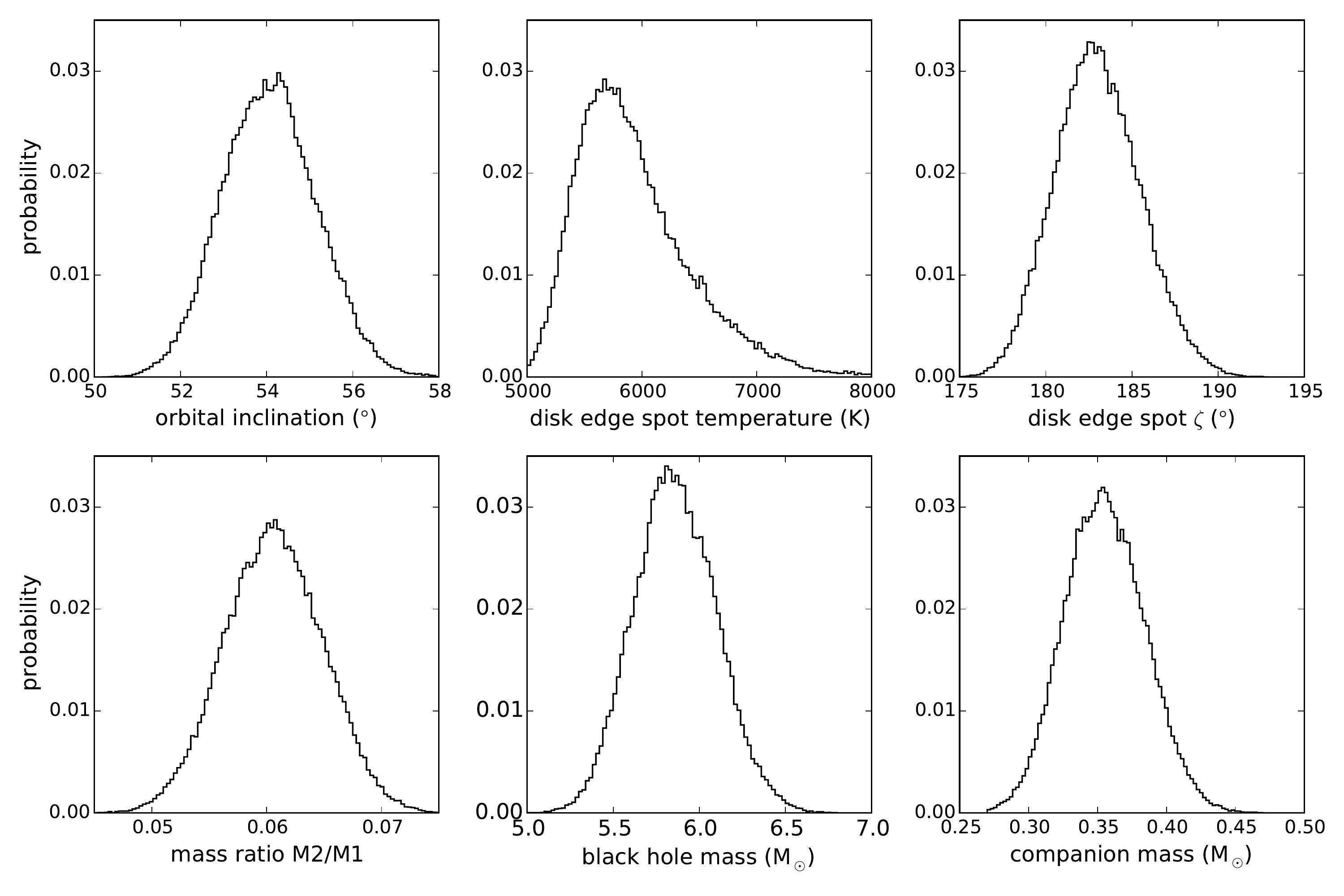}
  \caption{\emph{A posteriori} results for the basic model (Table~\ref{table:parameters}). Probability density distributions of the most important model parameters and the inferred component masses}
 \label{fig:2}
 \end{figure*}

\begin{table*}
\begin{minipage}[]{140mm}
  \caption{Light curve fitting results}
  \label{table:results}
  \begin{tabular}{l*{5}{c}}
  \hline
  								 &$V + I + H$ 			&$V + I + H$		 	&$V$ 				&$I$ 	 						&$H$ \\
  \hline
  $i$ 								&$54\fdg4\pm1\fdg1$	&$54\fdg1\pm1\fdg1$	&$58\fdg2\pm1\fdg9$	&$53\fdg6\pm1\fdg6$			&$50\fdg5\pm2\fdg2$  \\[5pt]
  $q$ 							&$0.0607\pm0.0042$	&$0.0606\pm0.0043$	&$0.0605\pm0.0044$ 	&$0.0602\pm0.0043$		 	&$0.0602\pm 0.043$  \\[5pt]
  $r_\mathrm{out}$ 					&$0.15^{+0.06}_{-0.02}$	&$0.14^{+0.05}_{-0.02}$	&$0.17^{+0.07}_{-0.04}$ 	&$0.16^{+0.08}_{-0.03}$ 			& $0.17^{+0.12}_{-0.04}$ \\[5pt]
  $h_\mathrm{out}$ 					&$0.23\pm0.12$		&$0.22\pm0.11$		&$0.26^{+0.38}_{-0.16}$	&$0.09^{+0.22}_{-0.06}$ 			& $0.09^{+0.23}_{-0.06}$ \\[5pt]
  $T_\mathrm{out} (10^3\mathrm{K})$ 	&$1.19^{+1.64}_{-0.69}$	&$1.29^{+2.00}_{-0.74}$	&$1.9^{+1.3}_{-1.2}$ 	&$2.0\pm1.2$		 			& $1.39^{+1.24}_{-0.79}$ \\[5pt]
  $T_\mathrm{edge} (10^3\mathrm{K})$	&$4.77\pm0.25$		&$4.83\pm0.27$		&$4.5^{+1.2}_{-0.7}$		&$5.3^{+3.2}_{-1.5}$ 			&$4.7^{+3.8}_{-2.0}$ \\[5pt]
  $\zeta_\mathrm{spot}$ 				&$182\fdg3\pm2\fdg0$ 	&$182\fdg9\pm2\fdg5$ 	&$180\fdg5\pm2\fdg1$	&$186\fdg4\pm3\fdg9$			&$159\fdg4\pm12\fdg0$ \\[5pt]
  $\Delta\zeta$ 						&$93\degr^{+65}_{-47}$	&$111\degr\pm55\degr$	&$49\degr^{+45}_{-27}$	&$91\degr^{+70}_{-54}$			&$71\degr^{+71}_{-51}$ \\[5pt] 
  $T_\mathrm{spot} (10^3\mathrm{K}$) 	&$5.8^{+0.6}_{-0.4}$ 	&$5.8^{+0.6}_{-0.4}$ 	&$5.9^{+2.1}_{-1.2}$		&$7.5^{+4.7}_{-2.6}$				&$10.4^{+6.7}_{-5.5}$ \\[5pt]
  \hline
  \multicolumn{2}{l}{Derived quantities}\\[5pt]
  $M_\mathrm{1}$ (M$_{\sun}$) 		&$5.88\pm0.23$   		&$5.86\pm0.24$   		&$5.06\pm0.31$		&$5.95\pm0.39$				&$6.75\pm0.65$ \\[5pt]
  $M_\mathrm{2}$ (M$_{\sun}$) 		&$0.35\pm0.03$		&$0.34\pm0.03$		&$0.31\pm0.03$		&$0.36\pm0.03$				&$0.41\pm0.05$\\ [5pt]
  \hline
  $\chi^2/\mathrm{d.o.f.}$				&2908/2325			&2314/2325			&1354/742			&$743/721$					&$791/844$ \\
  \hline				
  \end{tabular}

The numbers represent the median value of the parameters (not necessarily the values of the maximum likelihood fit) and their 68\% statistical uncertainty interval. The first column shows the result of fitting the data using the basic model as described in Section~\ref{sec:modeling}. The second column shows the effect of increasing the $V$-band errors by a factor of 1.36 to make the overall $\chi_\nu^2 \approx1$. Columns 3 ,4 and 5 show the result of fitting the light curves for each color individually.  
  The corresponding values for $i$ in CBO10 are\quad $V$: $51\fdg75\pm1\fdg05$,\quad $I$: $50\fdg13\pm{1\fdg35}$,\quad $H$: $51\fdg58\pm{3\fdg0}$.
  
\end{minipage}
\end{table*}  

Figure~\ref{fig:1} shows the best-fitting model light curves and their residuals with respect to the data. Figure~\ref{fig:2} shows the \emph{a posteriori} probability distributions of $i$, $T_\mathrm{spot}$, $\zeta_\mathrm{spot}$ and $q$, as well as those of the component masses. 
The probability distribution of the mass ratio is almost equal to its prior (well within the uncertainty of the latter), implying that the photometric data yield no further constraint over the spectroscopic value.

We obtain the probability distributions of the  mass of the binary components by calculating the latter for each instance of the model parameters. They are computed from the mass function 
\begin{equation}
\label{eq:mass function}
f(M)=\frac{K_2^3P}{2\pi G}=\frac{M_\mathrm{BH} \sin^3i}{(1+q)^2}
\end{equation}
where we use $K_2 = 435.4$~km\,s$^{-1}$ for the semi-amplitude of the radial velocity curve of the companion star \citep{2008MNRAS.384..849N}.

\begin{figure*}
  \includegraphics[width = 160mm]{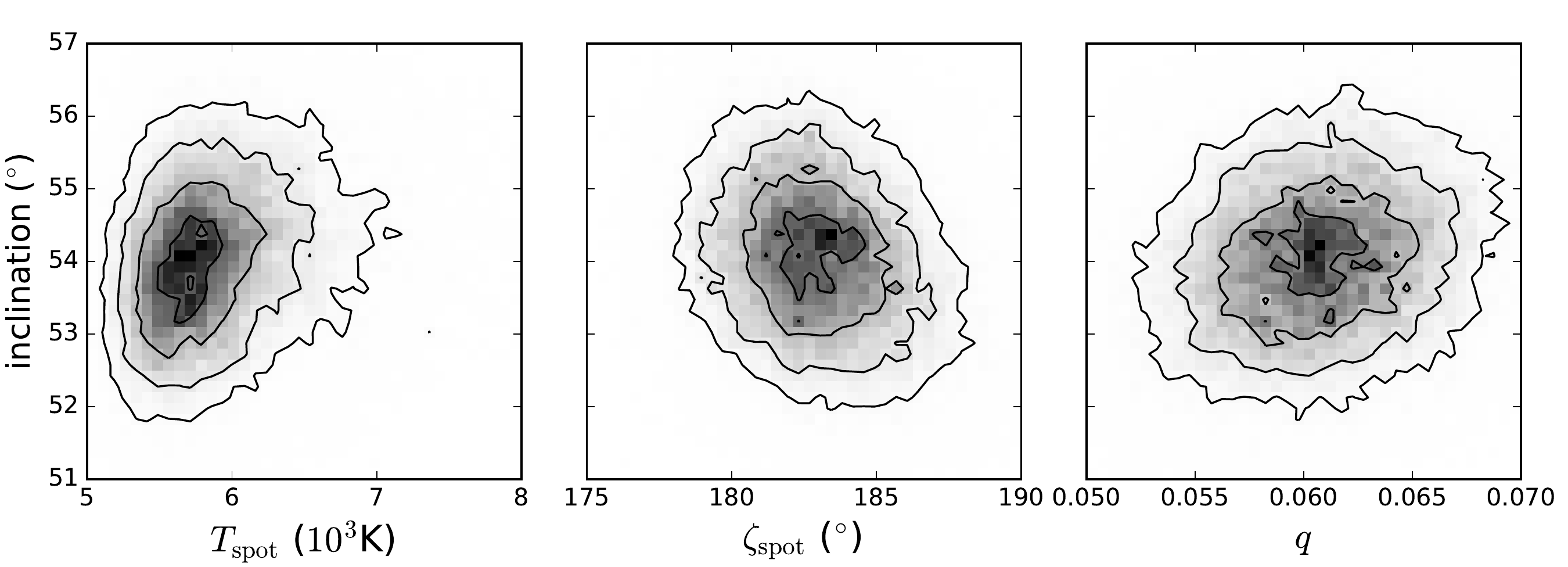}
  \caption{2d histograms for the MCMC run using all three filter bands, showing the covariance between the orbital inclination and three parameters of the model. Contours enclose approximately 17, 39, 68 and 86 per cent of the samples (corresponding to 0.5, 1.0, 1.5 and 2.0 sigma for the 1-d projections in the case of 2d Gaussian distributions). $T_\mathrm{spot}$ and $i$ are positively correlated, as higher disc flux makes the observed fractional ellipsoidal variations smaller than the intrinsic ones. The location of the disc spot $\zeta$ is well-determined, independent of $i$. The mass ratio $q$ and $i$ are only very slightly interdependent, as expected given the low value of $q$. 
 \label{fig:3}} 
 \end{figure*}

Table~\ref{table:results} shows the median values of all parameters and their 68\% uncertainty ranges as derived from the MCMC \emph{a posteriori} probability distributions. The best fit, using the basic model (first column), has a $V$-band reduced $\chi^2$ of about 1.8, and close to 1 for the other bands.  The excess variance in $V$ is possibly caused by (disc-) flickering, or it may indicate underestimation of the photometric errors by a factor of up to 1.36.
The values for the orbital inclination and the azimuth of the disc spot are well constrained and almost completely independent of the other parameters, whereas for all other parameters a wide range of values is compatible with the data. Figure~\ref{fig:3} shows that there is only weak covariance between the orbital inclination and the best-constrained model parameters. The influence of the disc fractions on the inclination outcome is discussed in Section~\ref{sec:df}.

The 3-band fit yields an inclination value of $54\fdg4\pm{1\fdg1}$ (see the first column in Table~\ref{table:results}).
Adjusting the $V$-band photometric uncertainties by a factor of 1.36 in order to obtain a reduced $\chi^2$ approximately equal to 1 in each band decreases the inclination to  $54\fdg1\pm{1\fdg1}$. The corresponding BH mass is $5.9\pm0.2$\,M$_{\sun}$, with an $0.34\pm 0.03$\,M$_{\sun}$ companion (column 2 in Table~\ref{table:results}). We adopt the latter values as our main result.

We also fit each light curve separately. The results are also shown in Table~\ref{table:results}. The best-fitting model parameter sets differ significantly between the single band light curves. The inclination value for $V$ is significantly higher than that obtained from fitting the 3 light curves together and the value for $H$ is lower. The larger photometric uncertainty and the low disc fraction in $H$ leave the disc parameters poorly constrained.

\section{DISCUSSION}
\label{sec:discussion}

\begin{figure}
\includegraphics[width = 80mm]{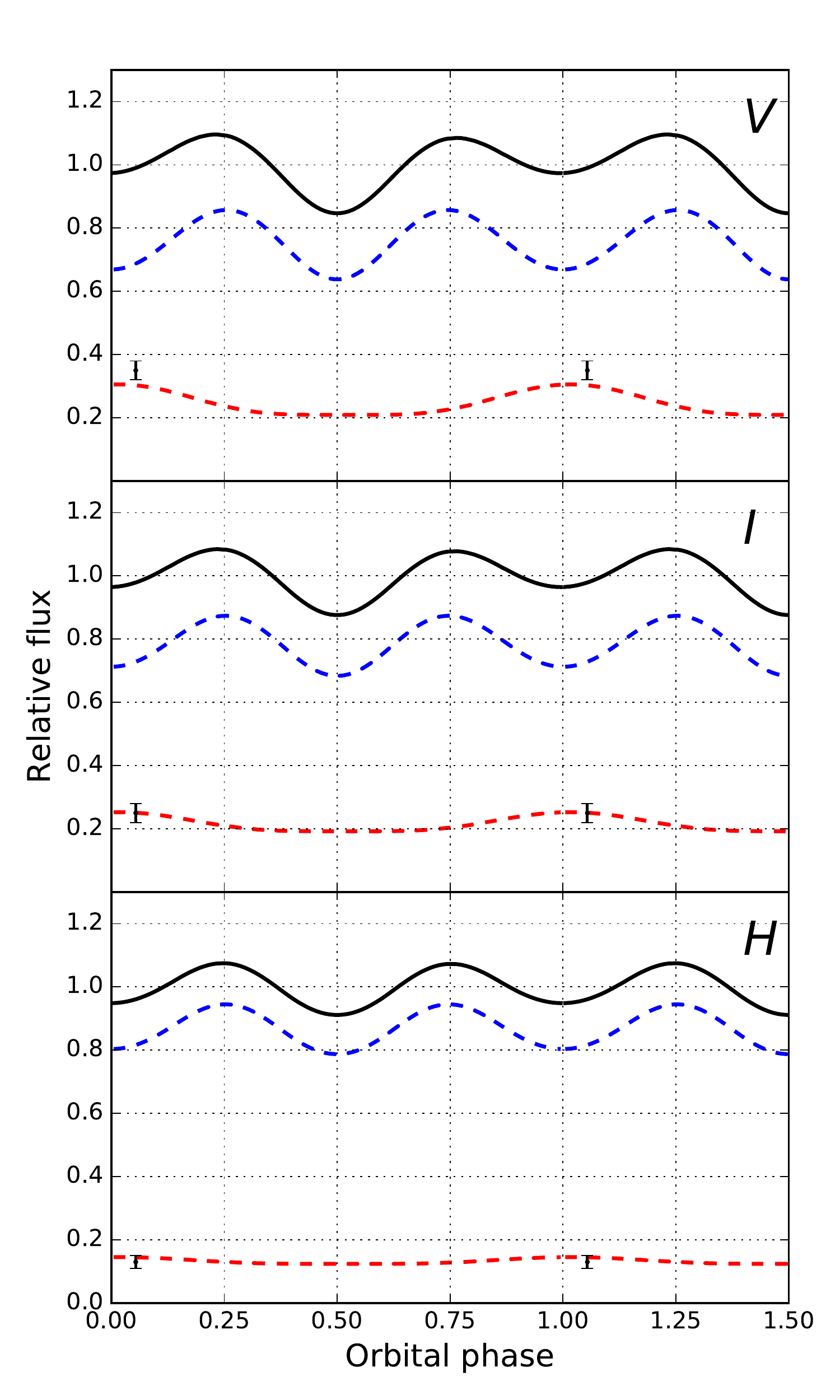}
  \caption{Components of the best-fitting model light curves. Top line: total flux, top dashed line: stellar flux, bottom dashed line: disc flux. The fluxes were normalized such that the orbital average of the total flux is equal to 1.0, corresponding to flux densities of $6.59\times10^{27},\,5.64\times10^{27}$ and $1.80\times10^{27}$ erg\,s$^{-1}$\AA$^{-1}$sr$^{-1}$ for $V$,$I$ and $H$, respectively. The errorbars represent the spectroscopically determined disc fractions of the total flux.}
  \label{fig:4}
\end{figure}

\begin{figure*}
\includegraphics[width = 140mm]{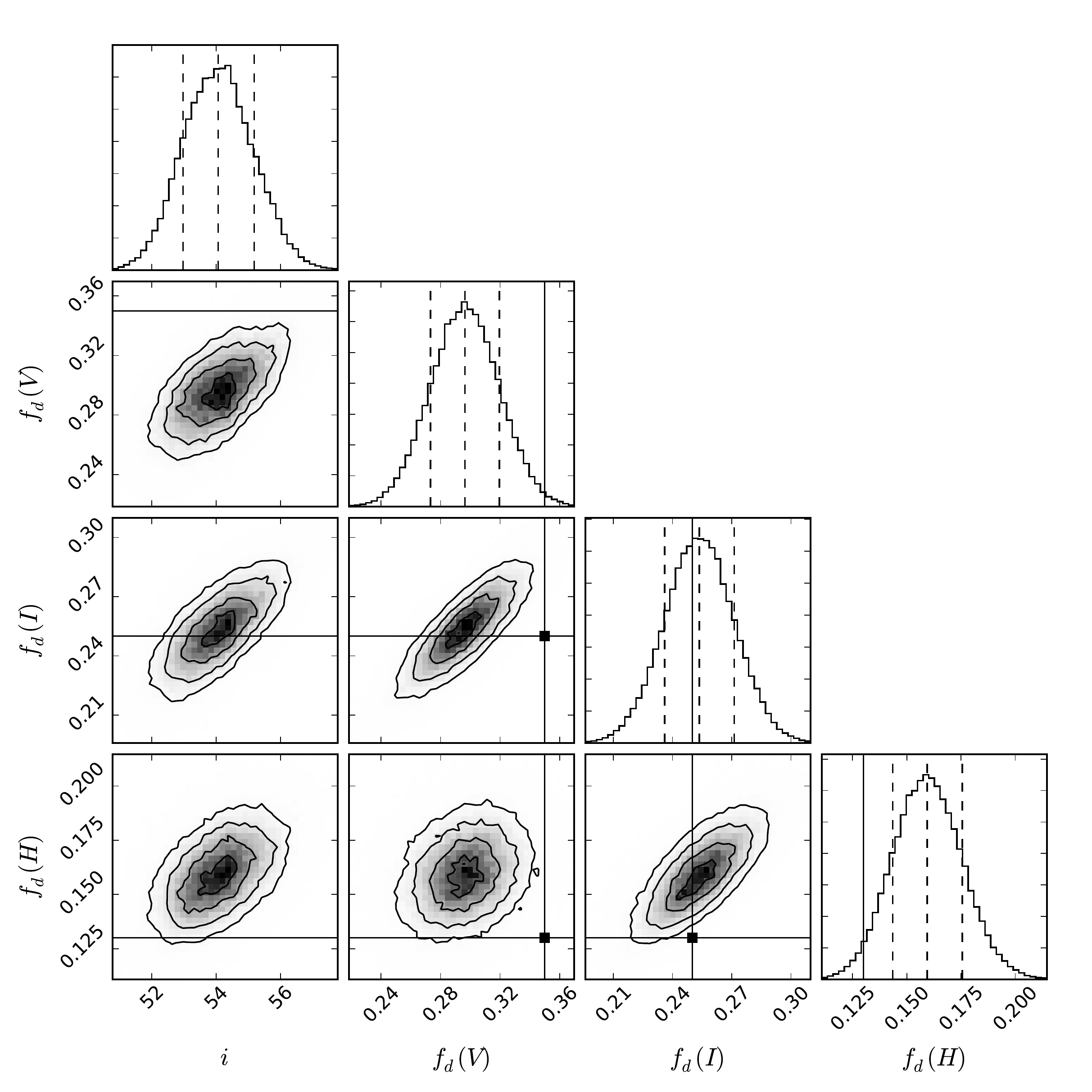}
  \caption{Covariance plots between the orbital inclination and the $V, I$ and $H$ posterior disc fractions (at orbital phase 0.054) in the first column, and  between the disc fractions in the rest of the figure. These are from the probability distributions resulting from the basic 3-band fit. Contours enclose approximately 17, 39, 68 and 86 per cent of the samples (corresponding to 0.5, 1.0, 1.5 and 2.0 sigma for the 1-d projections in the case of 2d Gaussian distributions). The dashed lines represent the median values and the 68\% most probable interval of each variable. The solid lines indicate the spectroscopically observed values of the latter. The distributions of the $V$- and $H$-band deviate signficantly from the spectroscopically determined values.}
  \label{fig:5}
\end{figure*}

In this paper we fit model light curves to SMARTS photometric light curves of A0620 in order to assess whether there are systematic errors associated with the light curve fitting due to the use of different software packages. Furthermore, we report and favour the results from a simultaneous fit to the 3 filter light curves.

In order to exclude the possibility of systematic error resulting from fixing the value of some parameters, we perform a number of tests. All tests are 3-band fits using the adjusted $V$-band photometric uncertainties. Lowering the donor $T_\mathrm{eff}$ from 4600\,K to 4200\,K (variant model V1) increases the median inclination from $54\fdg1\pm1\fdg1$ to $54\fdg4\pm1\fdg1$. Changing $r_\mathrm{in}$ from $0.1$ to $0.02$ (V2) and  $\gamma$ from $-0.75$ to $0.0$ (isothermal disc, V3) each result in a decrease of the median $i$ of less than $0\fdg05$.
For the gravity darkening coefficient $4\beta$ we used the value of 0.415 for all fits with $T_\mathrm{eff}=4600$\,K. Since this quantity is not very precisely known, we test the sensitivity of our results by repeating the fit with $4\beta=0.400$ (V4).  This changes the median $i$ from $54\fdg1$ to $54\fdg4$. Setting the edge temperature equal to the disc temperature at $r_\mathrm{out}$ (V5) does not change $i$ at all.\par
We also test for possible discretization error by repeating the 3-band fitting while doubling the number of disc and star tiles to 20,000 each. Finally, the accuracy of model light curve interpolation is tested by doubling the number of phases at which the model is computed. In both these tests the differences in the fitted parameters and the derived median BH mass are negligible. We conclude that none of the tests above shows evidence of significant systematic error in the inclination determination.

Although perhaps not immediately obvious in Figure~\ref{fig:1}, the best 3-filter band fit has residuals that are non-random, mainly in $V$ and $I$, in particular around the deepest minimum, giving the impression of a slight mismatch between the spectroscopic and photometric $T_0$. The individual fits to the same datasets in CBO10 (fig. 2, light curves V6, I6 and H6) show an even more obvious shift-like pattern of significantly non-random residuals. These authors tentatively ascribe these residuals to unresolved, phase-dependent flaring, which would have to be consistent over 4 years of passive state observations. Repeating our fit (weights adjusted) with a phase offset as an additional free parameter (V6) yields a lower $\chi^2$ of 2265 i.s.o. 2314 (for 2324 degrees of freedom), $\Delta \chi^2 = -49$, with an apparent offset of $0.0098\pm0.0014\,P$, or $240\pm40$\,s. It also increases the median inclination result by $0\fdg4$.\par
We consider this to be a real physical effect. Inaccuracy of the spectroscopic ephemeris used for folding the data can be ruled out as a cause for the apparent phase offset since the binary period is known to great precision and the $T_0$ value falls within the time frame of the SMARTS observations used here. The orbital period of A0620 is known to decay relatively fast \citep*{2014MNRAS.438L..21G}, but the value of $\dot{P}=-0.60\pm 0.08$\,ms\,yr$^{-1}$ is insignificant in the current context.

Several other causes of light curve distortion are possible.  An obvious one is that the accretion disc may not be axially symmetric.  Numerical models often produce asymmetric discs (eg, \citealt{2007MNRAS.378..785S}). There may be star spots on the secondary star, see e.g. \cite{2001MNRAS.326.1489L}, who find large cool starspots in the RS CVn binary XY UMa using eclipse mapping. These effects could cause significant, wavelength dependent, distortions of the light curve, including displacing the phase of photometric conjunction from that of spectroscopic conjunction.
Asymmetry of the stellar flux, could also cause a minor phase-dependent distortion of the radial velocity measurements. We conclude that both the CBO10 model and ours do not fully explain the light curves, which leaves the probability of a small systematic error in the determination of $i$. Since the data do not allow a resolution of this issue, we favour the zero-shift value of $54\fdg1\pm1\fdg1$ for the orbital inclination, with masses $5.86\pm0.24M_{\sun}$ and $0.34\pm0.03M_{\sun}$ for the BH and companion star, respectively. The CBO10 values are $i=51\fdg0\pm0\fdg9$ and $M_\mathrm{BH}=6.61\pm0.25M_\odot$.

While CBO10 fit the $V$, $I$, and $H$ light curves individually, they find inclination values that are mutually compatible: $51\fdg75\pm1\fdg05$, $50\fdg13\pm{1\fdg35}$ and $51\fdg58\pm{3\fdg0}$, respectively. 
When we do the same, allowing the model parameters to be different for each bandpass, like CBO10 do, we find $58\fdg2\pm1\fdg9$, $53\fdg6\pm1\fdg6$ and $50\fdg5\pm2\fdg2$. In this case the disc fractions of the models at phase 0.054 are: $0.35\pm0.03, 0.25\pm 0.03$ and $0.13\pm0.02$, essentially identical to the measured values used as inputs. The values for the 3-bandpass model fit are $0.30\pm0.02, 0.25\pm 0.02$ and $0.16\pm 0.02$. The difference may be evidence of a systematic difference between the model of CBO10 and ours. The role of the disc fractions is further discussed in Section \ref{sec:df}.

\subsection{Disc fractions}  
\label{sec:df}
The spectroscopically determined fractional contributions of non-stellar light (disc plus spot) to the total flux are essential for determining the orbital inclination of the binary. Figure~\ref{fig:4} shows the stellar and disc flux of the best fitting light curves for the basic model, as well as the spectroscopic disc fractions. The disc component has its maximum just after phase 0, partly filling in the secondary minimum of the ellipsoidal variation of the companion star. In our model this implies that the disc spot center trails the companion star slightly, by $3\degr\pm2\degr$. This conflicts with the usual view that it should lead the companion if it is caused by the accretion flow impinging on the disc. It must be noted that the disc fractions at phase 0.054 are not representative of the disc fractions over the full orbit. Phase 0.054 almost coincides with the highest contribution of the spot to the non-stellar flux. In $V$ for instance, it is responsible for about 10\% of the total flux, while it contributes less than 1.5\% of the total flux in the phase range 0.3--0.7.\\
Figure~\ref{fig:5} shows the joint posterior probability distributions of the inclination and the $V$-, $I$- and $H$-band disc fractions at phase 0.054 resulting from our fit to the 3-filter band light curves simultaneously. There is a strong covariance between the disc fractions and the inclination, as expected. The larger the fractional contribution of the disc to the total flux, the lower the fractional amplitude of the ellipsoidal variations, given a constant intrinsic ellipsoidal light curve. Conversely, when modeling the observed light curves, the intrinsic amplitude must increase when the disc fraction is higher, resulting in a higher inclination. The single-passband fits show that a good fit can be obtained in each band that satisfies the disc fraction data, while yielding significantly different inclination values (see Table~\ref{table:results}). The 3-passband fit yields an $I$-band \emph{a posteriori} disc fraction close to its data value, while those in $V$ and $H$ differ by $\approx2\sigma$ from theirs. Changing the disc temperature profile (see Section~\ref{sec:discussion}) has no effect on this apparent difference.\\
Perhaps the uncertainties quoted by CBO10 are too small. We note that the disc fractions were all derived more indirectly, least so for $V$, more for $H$, and even more for $I$.  In order to test the impact of larger uncertainties we performed a fitting run, arbitrarily multiplying them by 1.0, 1.5 and 2.0 for $V$, $I$, and $H$, respectively. This increases the median inclination to $54\fdg8\pm1\fdg3$, with \emph{a posteriori} disc fractions 0.31, 0.27 and 0.18.\\
It is also possible that part of the non-stellar flux is contributed by a jet producing a near infrared excess. With an 8.46\,GHz flux density in quiescence  of $\approx 50\,\mu$Jy \citep{2006MNRAS.370.1351G}, an inverted spectrum and/or a spectral break above $2x10^{14}$\,Hz is required for the jet to contribute significantly to the $H$-band flux. The higher disc fraction in the $V$ band would then be produced by a smaller, hotter disc. In either case more accurate BH mass determinations will benefit from obtaining time-resolved spectra simultaneously with photometry, ideally continuously, in order to eliminate the effects of short-term fluctuations in the disc contribution to the total flux.

\section{CONCLUSIONS}
Using our model and modeling software, fitting $V$, $I$ and $H$ data simultaneously, we find a value of  $54\fdg1\pm1\fdg1$ for the orbital inclination of A0620, with a BH mass of $5.86\pm0.24M_{\sun}$. The results of the tested variant models differ from those of the basic by only a fraction of the statistical uncertainties.\\
Although it yields a good overall fit to the three observed light curves, the result of our 3-band simulation is at odds with the spectroscopic disc fractions. We have no conclusive evidence as to why this is the case.\\
We discovered an apparent offset between the spectroscopic and photometric ephemeris of A0620. At this time we have no explanation for this.\\
The conclusion seems justified that the mass determination of this BH system still suffers from systematic uncertainties which are larger than the statistical uncertainty.
As it is known that non-stellar flux in low-mass X-ray binaries fluctuates on short timescales, obtaining time-resolved spectra simultaneously with photometry will enable to better separate the stellar and non-stellar contributions to the observed flux, and therefore improve the accuracy of dynamical BH mass measurements.

\section{AKNOWLEDGMENTS}
The authors thank Charles Bailyn (Astronomy Department, Yale University) for kindly providing the YALO and SMARTS photometric data.
PGJ acknowledges support from European Research Council Consolidator Grant 647208.

\begin{bibliography}{A0620-00_submit}
\bibliographystyle{mnras}
\end{bibliography}

\end{document}